# New Photometric Observations and the 2015 Eclipse of the Symbiotic Nova Candidate ASAS J174600-2321.3

**Franz-Josef Hambsch**
*Oude Bleken 12, B-2400 Mol, Belgium; American Association of Variable Star Observers (AAVSO), Cambridge, MA; Bundesdeutsche Arbeitsgemeinschaft für Veränderliche Sterne e.V. (BAV), Berlin, Germany; and Vereniging Voor Sterrenkunde (VVS), Brugge, Belgium; hambsch@telenet.be*

**Stefan Hümmerich**
*Stiftstr. 4, Braubach, D-56338, Germany; American Association of Variable Star Observers (AAVSO), Cambridge, MA; and Bundesdeutsche Arbeitsgemeinschaft für Veränderliche Sterne e.V. (BAV), Berlin, Germany; stefan.Hümmerich@gmail.com*

**Klaus Bernhard**
*Kafkaweg 5, Linz, 4030, Austria; American Association of Variable Star Observers (AAVSO), Cambridge, MA; and Bundesdeutsche Arbeitsgemeinschaft für Veränderliche Sterne e.V. (BAV), Berlin, Germany*

**Sebastián Otero**
*Olazabal 3650-8 C, Buenos Aires, 1430, Argentina; American Association of Variable Star Observers (AAVSO), Cambridge, MA*

**Abstract** The eclipsing binary system ASAS J174600-2321.3, which has shown a conspicuous brightening of ~4 magnitudes (V) in the past, was recently identified as a symbiotic nova candidate. A long-term photometric monitoring program was initiated in July 2014. In its present active stage, the system shows deep eclipses with an amplitude of ~3.5 magnitudes (V) that occur about every 33 months. In order to monitor the eclipse of 2015, *AAVSO Alert Notice 510* was issued. During the ensuing campaign, AAVSO observers obtained 338 measurements in Johnson B, 393 measurements in Johnson V, and 369 measurements in Cousins I, as well as 27 visual observations. The present paper presents and analyzes these data from the AAVSO International Database, along with observations from the aforementioned photometric monitoring program. From these data, we were able to refine the orbital period to $P_{orb}$ = 1012.4 days. Furthermore, the data are suggestive of a slight decrease in mean brightness, which—if proven real—might indicate a decline of the outburst.

## 1. Introduction

ASAS J174600-2321.3 = EROS2-cg1131n13463 = 2MASS J17460018-2321163, situated in Sagittarius at R. A. (J2000) = $17^h\ 46^m\ 00.180^s$, Dec. –23° 21' 16.37" (UCAC4; Zacharias *et al.* 2012), is a long-period eclipsing binary system ($P_{orb}$ ~ 33 months). The primary star exhibits the spectral characteristics of a reddened, early F-type supergiant and is likely a white dwarf (WD) currently in outburst. The secondary component is a giant of spectral type late M. The stellar parameters, as derived in Hümmerich *et al.* (2015a; hereafter Paper 1) are given in Table 1. (For more information on the system components, the reader is referred to Paper 1, especially section 3.) The system's photometric variability is characterized by orbital variability (eclipses), semiregular pulsations, and a conspicuous brightening of ~4 magnitudes (V) in the recent past, which has been well documented in available sky survey data (Figure 3 and Paper 1, Figure 1). The outburst has already lasted for more than ~12.8 years and continues to the present day.

From the observed characteristics, ASAS J174600-2321.3 was identified as a symbiotic nova candidate (Paper 1). A long-term photometric monitoring program was initiated. Additionally, *AAVSO Alert Notice 510* (AAN 510) was issued in order to get good multicolor coverage of the 2015 eclipse (Hümmerich *et al.* 2015b). The present paper presents and discusses the observations from the aforementioned sources. It is organized as follows: section 2 presents past and present observations of our target, which are analyzed and discussed in section 3. We conclude in section 4.

## 2. Observations

Before considering the photometric data, it is noteworthy to point out the considerable line-of-sight extinction to ASAS J174600-2321.3. Based on the calculations of Schlafly and Finkbeiner (2011), who employ the colors of stars with spectra in the Sloan Digital Sky Survey and measure reddening as the difference between the measured and predicted colors of a star, we estimate an interstellar extinction of $A_V$ ≈ 2.4 mag.

Table 1. Stellar parameters and binary separation of ASAS J174600-2321.3, as derived in Paper 1.

| Parameter | Value |
|---|---|
| Stellar Parameters | $M_g$ ≈ 1.5 $M_\odot$ |
| (Red Giant) | $R_g$ ≈ 145 $R_\odot$ |
| | $T_{eff}$ ≈ 3,130 K |
| | Spectral type ~M7 |
| (White Dwarf) | $M_{wd}$ ≈ 0.5 $M_\odot$ |
| Binary Separation | R ≈ 2.5 AU |

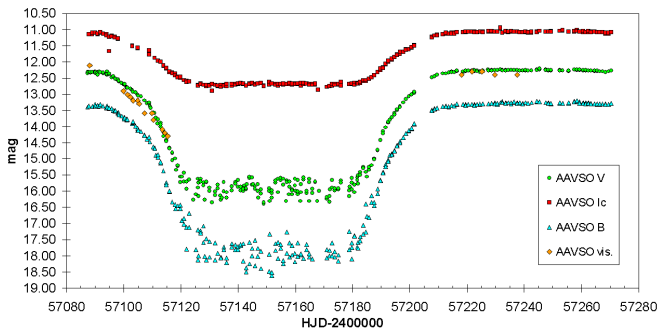

Figure 1. Light curve of ASAS J174600-2321.3, based on data obtained by AAVSO observers during the campaign initiated via *AAVSO Alert Notice 510*. Obvious outliers have been removed by visual inspection.

Table 2. Observers of ASAS J174600-2321.3.

| Name | AAVSO Observer Initials | Number of Observations |
|---|---|---|
| Salvador Aguirre | ASA | 1 |
| Teofilo Arranz | ATE | 7 |
| Alejandra Arranz Lázaro | AALB | 2 |
| Lewis Cook | COO | 16 |
| Brian Cudnik | CKB | 3 |
| Shawn Dvorak | DKS | 15 |
| Franz-Josef Hambsch | HMB | 912 |
| Michael Heald | HMH | 36 |
| Gordon Myers | MGW | 70 |
| Peter Robert Nelson | NLX | 23 |
| Steve O'Connor | OCN | 3 |
| Andrew Pearce | PEX | 14 |
| Larry Shotter | SLH | 8 |
| Peter John Starr | SPET | 18 |
| Robert R. Young | YON | 1 |

and E(B–V) ≈ 0.78 mag. If not indicated otherwise, all light curves illustrated in this and the following sections are based on non-corrected data.

2.1. AAVSO observing campaign

During the observing campaign initiated by *AAN 510*, AAVSO observers obtained 338 measurements in Johnson B, 393 measurements in Johnson V, and 369 measurements in Cousins I ($I_C$), as well as 27 visual observations (Kafka 2015). A light curve based on these data is shown in Figure 1. In Table 2 we gratefully acknowledge the efforts of the contributing observers.

As the agreement for B and V data between AAVSO observers is very good, we chose to combine the corresponding datasets for the following analyses; some obvious outliers were removed by visual inspection. For $I_C$ data, the situation is more complicated, though. There are significant mean magnitude differences (up to ~0.7 mag.) between several observers' datasets. An investigation of the respective datasets indicates that, most likely, the use of different comparison star sets is responsible for the observed discrepancies. As a solution to this problem, we have chosen to base our analyses on the most comprehensive and homogeneous $I_C$ dataset available. Because of the availability of extensive datasets of CCD photometry, visual observations were not included into the analyses in section 3.

2.2. Photometric monitoring program

In order to study the long-term photometric behavior of the system, a monitoring program was initiated at the Remote Observatory Atacama Desert (ROAD; Hambsch 2012). Starting on July 17, 2014, observations have been taken with an Orion Optics, UK Optimized Dall Kirkham 406/6.8 telescope and a FLI 16803 CCD camera. Images were taken through Astrodon Photometric B, V, and $I_C$ filters. Initially, observations were carried out in V only; with the start of the AAVSO observing campaign, simultaneous observations in B and $I_C$ were added to the observing program. ROAD data taken during the AAVSO campaign (912 observations) are not designated separately but are included with the AAVSO data; they can be identified by the observer code "HMB". A light curve of the observations carried out prior to the start of the AAVSO campaign is shown in Figure 2. Two measurements per night were taken; the plot is based on nightly averages.

2.3. Sky survey data and long-term light curve

In our initial investigation of ASAS J174600-2321.3 (Paper 1), we procured data from the EROS-2 (Renault *et al.* 1998), ASAS-3 (Pojmański 2002), CMC14 (Evans *et al.* 2002), and APASS (Henden *et al.* 2012) archives. EROS-2 $B_E$ and $R_E$ magnitudes were transformed to V and $I_C$, respectively. Approximate $I_C$ values were derived from APASS r' and i' photometry. The CMC14 observation was transformed to V. Furthermore, obvious outliers with a photometric quality flag of D (= "worst data, probably useless") were removed from the ASAS-3 dataset. A detailed description of the various data sources and reduction processes involved can be found in Paper 1, section 2.1.

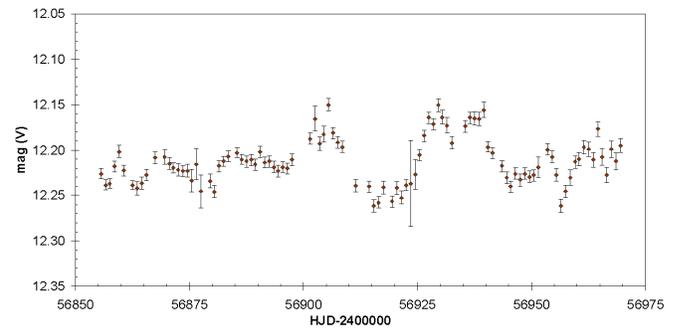

Figure 2. Light curve of ASAS J174600-2321.3, based on ROAD data, which have been acquired during the authors' photometric monitoring program. The plot is based on nightly averages (see text for details).

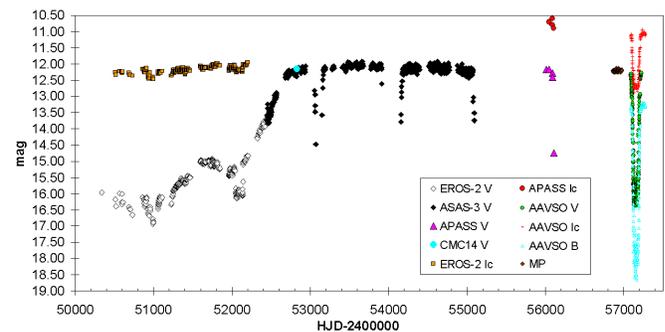

Figure 3. Light curve of ASAS J174600-2321.3, based on various data sources, as indicated in the legend on the right. "MP" designates data from the long-term monitoring program (see section 2.2). Obvious outliers have been removed from AAVSO data by visual inspection.

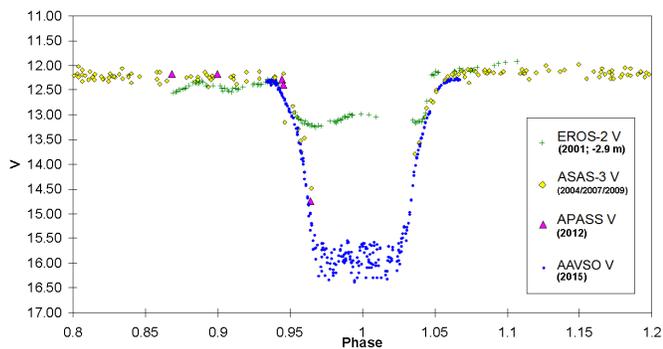

Figure 4. Phase plot of ASAS J174600-2321.3, illustrating all primary eclipses. The plot is based on various data sources, as indicated in the legend on the right, and folded with the elements given in Equation (1).

Table 3. Mean Brightness of ASAS J174600-2321.3 during the total phase of the 2015 eclipse and amplitude of the 2015 eclipse in B, V, and $I_C$.

| Mean Brightness During Totality Phase | Amplitude of the 2015 Eclipse |
|---|---|
| mean B = 17.9 mag. | $\Delta B \approx 4.6$ mag. |
| mean V = 15.8 mag. | $\Delta V \approx 3.6$ mag. |
| mean $I_C$ = 12.7 mag. | $\Delta I_C \approx 1.6$ mag. |

Figure 3 shows the long-term light curve of ASAS J174600-2321.3, which has been based on sky survey data, the authors' long-term monitoring program (designated as "MP" in the legend), and AAVSO campaign data.

## 3. Discussion

### 3.1. Orbital period

In order to fit the new observations, the previously determined orbital period of $P_{orb}$ = 1011.5 days (see Paper 1, section 3.1) had to be adjusted slightly. From an analysis of V-band data for all eclipses, the following improved orbital ephemeris was derived.

$$\text{HJD (Min I)} = 2456142 + 1012.4 \, (\pm 0.2) \times E \quad (1)$$

Figure 4 shows the resulting phase plot and illustrates all primary eclipses that have been observed so far. In order to facilitate comparison, EROS-2 V magnitudes have been shifted by the indicated amount to match V magnitudes from ASAS-3, APASS, and AAVSO.

### 3.2. Eclipse depth

Only rudimentary coverage of the eclipses existed after the onset of activity and the considerable brightening of the system, which has been shown to be restricted to the primary star of the system (the outbursting WD; see Paper 1, section 3.1). In particular, no observations were taken during the total phase of any ensuing eclipse. Thus, for the first time, AAVSO campaign data allow the measurement of the depth of the eclipse in outburst stage. Average values and amplitudes are given in Table 3.

As the seat of the outburst is proposed to be the WD primary component of ASAS J174600-2321.3, and the observed eclipse shows a pronounced phase of totality, in which the red giant passes in front of the WD, it might reasonably be assumed that the system's brightness returns to its pre-outburst value of ~16.5 mag (V) during the total phase of the eclipse. This, however, is obviously not the case—the system's brightness hovers around 15.8 mag. (V) during the total phase of the 2015 eclipse instead (see Table 3). This difference in mean brightness is also reflected in a bluer color index at this phase. Explanations for these interesting phenomena are discussed below in section 3.5.

### 3.3. Variability during total eclipse

Both components of ASAS J174600-2321.3 were shown to exhibit light variability, which is likely due to pulsations (see Paper 1, section 3.4). The observed semiregular pulsations of the red giant component dominated the light changes of the system before the proposed outburst and the corresponding rise in mean magnitude. They were most obvious in EROS-2 $I_C$ band data, exhibiting a mean amplitude of $\Delta I_C \approx 0.1$ mag. Although the pulsations were rather ill-defined throughout the covered timespan, a dominant signal at $P \approx 56$ days was identified (see Paper 1, Figure 14).

Light variability was observed in the form of a rebrightening around the time of mid-eclipse during the total phase of the 2001 eclipse—the only eclipse with good coverage during totality phase in the past. In regard to phase and amplitude, the observed variability was in agreement with the red giant's pulsations and was interpreted in this vein (see Paper 1, section 3.4).

We have searched for variability during the total phase of the 2015 eclipse. As the system gets considerably faint during the phase of totality (B ≈ 17.9 mag.; V ≈ 15.9 mag.; $I_C$ ≈ 12.7 mag.), the measurement errors increase markedly, too, leading to increased scatter during this part of the light curve. These measurement uncertainties are most pronounced in B and V band data, obliterating any trace of variability in B, and rendering detection in the V light curve a borderline case. However, the situation is different in $I_C$ data, which clearly show indications of variability (Figure 5); again, a central rebrightening is observed.

In terms of duration and amplitude, the observed light changes are in agreement with the pulsational variability of the red giant component. However, it seems a curious fact that the pulsations of the red giant should exhibit nearly identical maxima around the time of mid-eclipse during both the 2001 and 2015 totalities. Therefore, it seems worthwhile to consider alternative explanations for the observed phenomenon. Snyder and Lapham (2008) investigated eclipse brightening features in 18 binary systems. Eclipse brightening has been identified throughout a wide range of system configurations, from detached to over-contact binaries and short to very long period systems such as ε Aurigae. Brightenings around mid-eclipse have been described for several objects.

Snyder and Lapham (2008) examined gravitational lensing, microlensing and refracted electromagnetic radiation ("prism effect") as possible causes. In the case of the interacting symbiotic binaries with their complex light curves (Skopal 2008, for example) and often variable eclipse shapes (Bruch *et al.* 1994, for example), other scenarios involving mass transfer/mass accretion might be conceivable. However, at this point, we can only speculate on the nature of the observed brightening. Spectroscopic as well as photometric coverage of upcoming eclipses will be necessary to decide on the true nature of the observed phenomenon.

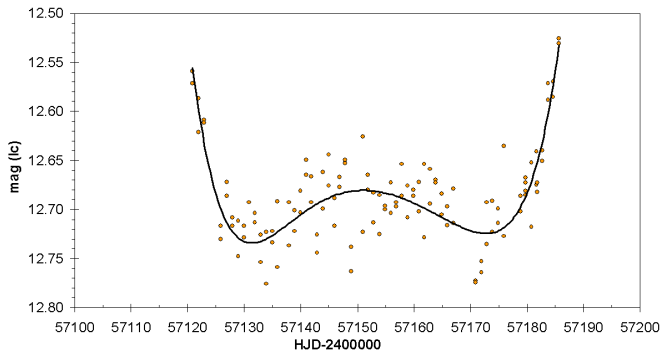

Figure 5. Detailed view of the total phase of the 2015 eclipse, based on AAVSO $I_C$ data. The solid line indicates a 6th order polynomial fit to the data.

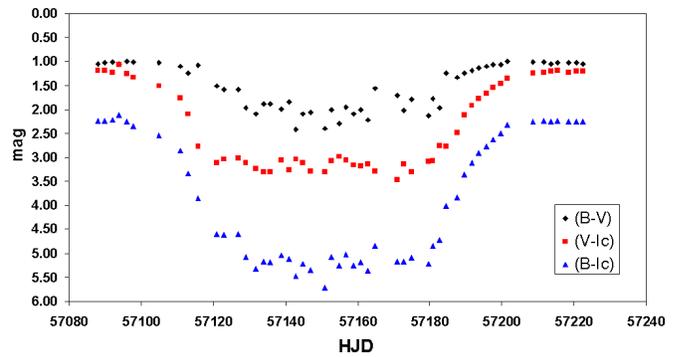

Figure 6. Color curves in (B–V), (V–$I_C$), and (B–$I_C$) of ASAS J174600-2321.3, based on two-point averages of nightly mean values of the data obtained by AAVSO observers during the campaign initiated via *AAVSO Alert Notice 510*. Obvious outliers have been removed by visual inspection.

3.4. Variability at maximum visual brightness

After the dramatic rise in brightness, which culminated with reaching ~12.2 mag. (V) at about HJD 2452600, the WD primary component of the system came to dominate the flux at optical wavelengths. Although the light from the WD now effectively swamped the red giant's pulsations in V, there were indications of pulsational-like variability in the ASAS-3 light curve at maximum visual light after the brightening event. The observed variability was rather irregular and poorly defined, although there is indication of a period of about 80–90 days, particularly around HJD 2453500 (see Paper 1, Figure 13).

After the end of ASAS-3 observations, there followed a gap in photometric coverage of about 2,000 days, interrupted only by a few APASS measurements around HJD 2456000. New data were made available with the start of the photometric monitoring program described in section 2.2. Interestingly, the data obtained paint quite a different picture of the variability at maximum visual brightness (Figure 2); the rather slow light changes detected in ASAS-3 data seem to have been substituted by more irregular variability on shorter time-scales. An analysis of the new data with PERIOD04 (Lenz and Breger 2005) and the CLEANest algorithm (Foster 1995) suggests the existence of a weak signal at P ≈ 33 days. However, because of the irregularity of the observed light changes and the short time baseline, the significance of this detection is doubtful. Still, it is intriguing to connect the apparently shortened time-scale of variability at maximum light to a decrease in diameter of the pseudophotosphere of the WD, which would go along well with the apparent slight decrease of the system's mean brightness in recent observations, which is commented on below (see section 3.6).

3.5. Color curves and indices

Average color indices and color curves were constructed from AAVSO campaign data. Because of the faintness of the system during the total phase of the eclipse, a lot of scatter is present during these parts of the light curve, which is attributable to a corresponding increase in measurement errors. In order to augment this situation, we have chosen to use two-point averages based on nightly mean values in the construction of the color curves, which are shown in Figure 6. Average dereddened color indices at maximum light and the middle of total eclipse are presented in Table 4, along with color indices derived from earlier observations in Paper 1.

While the observed dereddened color indices at maximum visual brightness are in accordance with the earlier observations ((B–V)$_0$ ≈ 0.2; see Paper 1, Table 4), there is considerable discrepancy between the measured (V–$I_C$) indices before the onset of activity in the system and during the total phase of the 2015 eclipse (Table 4).

With the red giant component dominating before the brightening event and during the total phase of the eclipse (and the WD's contribution assumed to be negligible during both phases), one would expect the colors to match more closely during these phases. As has been pointed out above, though, the system is also ~0.7 mag. (V) brighter during the total phase of the 2015 eclipse (~15.8 mag. (V)) than at quiescence (~16.5 mag. (V); see section 3.2).

Several mechanisms might contribute to the observed phenomena. For instance, the contribution of the red giant to the total flux of the system might be enhanced, possibly due to the heating of its outer layers by the intense ultraviolet radiation from the WD. On the other hand, there might be a significant flux contribution from the WD even at this phase; for example,

Table 4. Average dereddened color indices at mid-eclipse and maximum visual brightness, as derived from AAVSO campaign data and earlier observations.

| Mid-eclipse (2015 eclipse) | Maximum Visual Brightness | Derived from Earlier Observations (Paper 1) |
|---|---|---|
| (B–V)$_0$ ≈ 1.2 | (B–V)$_0$ ≈ 0.14 | (B–V)$_0$ ≈ 0.2 epoch: HJD 2456008.794 (max. vis. brightness) |
| (V–$I_C$)$_0$ ≈ 2.1 | (V–$I_C$)$_0$ ≈ 0.14 | (V–$I_C$)$_0$ ≈ 3.0 mean value at min. vis. brightness (HJD < 2451112.53) |
| (B–$I_C$)$_0$ ≈ 3.3 | (B–$I_C$)$_0$ ≈ 0.28 | n/a |

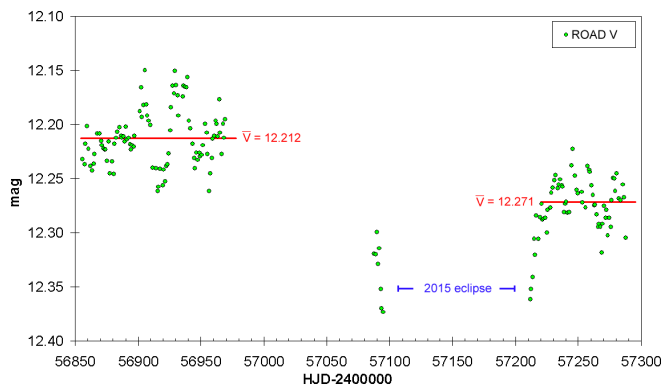

Figure 7. Detailed view of photometric observations taken at ROAD observatory, based on nightly mean values. The solid lines indicate the mean magnitudes of data taken before and after the 2015 eclipse.

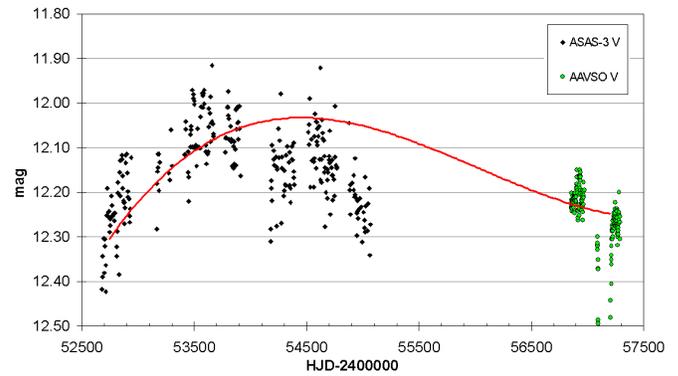

Figure 8. Light curve of ASAS J174600-2321.3, based on ASAS-3 and AAVSO V band data. The solid line represents a polynomial fit to the data.

the light of the WD might permeate the outer layers of the tenuous atmosphere of the red giant.

However, the most significant contribution will likely arise from the huge amount of circumstellar gas ionized by the intense radiation of the outbursting WD. As the volume occupied by the gas will likely exceed the diameter of the cool giant, there will be a significant contribution to the light output of the system even during the total phase of eclipse. This is in accordance with the above mentioned blueward evolution of the system's color indices, which has also been observed in other symbiotic novae (see, for example, Kenyon 1986).

3.6. Mean magnitude trend

Recent data are suggestive of a slight decrease in mean magnitude of the system, especially after the end of the 2015 eclipse. In order to verify its reality, we have investigated this phenomenon in ROAD data, which constitute the most extensive and homogeneous dataset available. No change in instrumental setup was performed at the ROAD observatory during the monitoring program and the AAVSO campaign. As shown in Figure 7, the observed decrease in mean magnitude seems to be real and significant.

We have combined ASAS-3 and AAVSO V-band data to search for overall mean magnitude trends. The data are suggestive of a long-term mean brightness trend, which we have visualized with a polynomial fit to the data in Figure 8. It is tempting to associate the decreasing optical flux output of the system with a possible shrinking of the WD's pseudophotosphere, which might indicate a decline of the outburst.

However, because of the considerable measurement errors of ASAS-3 data and the intrinsic brightness fluctuations of the system, further observations are needed to establish the reality of the proposed long-term mean magnitude trend.

4. Conclusion

We have presented and analyzed observations of the eclipsing symbiotic nova candidate ASAS J174600-2321.3 from our photometric long-term monitoring program and the recent AAVSO observing campaign of the 2015 eclipse, which was initiated via *AAVSO Alert Notice 510*. AAVSO observers obtained 338 measurements in Johnson B, 393 measurements in Johnson V, and 369 measurements in Cousins I, as well as 27 visual observations. These data allowed the measurement of the eclipse depth in outburst stage for the first time ($\Delta B \approx$ 4.6 mag; $\Delta V \approx$ 3.6 mag; $\Delta I_C \approx$ 1.6 mag). By combining all available V-band data, the orbital period could be refined to $P_{orb}$ = 1012.4 days. The system is bluer and brighter during the totality phase than before the beginning of the outburst phase and the corresponding rise in magnitude, which is mostly likely due to a significant light contribution by ionized circumstellar gas. Furthermore, recent data are suggestive of a slight decrease in mean brightness, which – if proven real – might indicate a decline of the outburst. Further photometric and spectroscopic studies of this interesting object are encouraged.

5. Acknowledgements

We acknowledge with thanks the variable star observations from the AAVSO International Database contributed by observers worldwide, which have made this research possible. We would also like to thank the anonymous referee for suggestions that helped to improve the paper. This work has made use of the SIMBAD and VizieR databases operated at the Centre de Données Astronomiques (Strasbourg) in France. This research has also made use of EROS-2 data, which were kindly provided by the EROS collaboration. The EROS (Expérience pour la Recherche d'Objets Sombres) project was funded by the CEA and the IN2P3 and INSU CNRS institutes. Furthermore, this research has employed data products from the Two Micron All Sky Survey, which is a joint project of the University of Massachusetts and the Infrared Processing and Analysis Center/California Institute of Technology, funded by the National Aeronautics and Space Administration and the National Science Foundation.

References

Bruch, A., Niehues, M., and Jones, A. F. 1994, *Astron. Astrophys.*, **287**, 829.
Evans, D. W., Irwin, M. J., and Helmer, L. 2002, *Astron. Astrophys.*, **395**, 347.
Foster, G. 1995, *Astron. J.*, **109**, 1889.

Hambsch, F.-J. 2012, *J. Amer. Assoc. Var. Star Obs.*, **40**, 1003.
Henden, A. A., *et al.* 2012, Data Release 3 of the AAVSO All-Sky Photometric Survey (http://www.aavso.org/apass).
Hümmerich, S., Otero, S., Tisserand, P., and Bernhard, K. 2015a, *J. Amer. Assoc. Var. Star Obs.*, **43**, 14.
Hümmerich, S., Otero, S., Tisserand, P., Bernhard, K., and Templeton, M. R. 2015b, *AAVSO Alert Notice 510*, 1.
Kafka, S. 2015, observations from the AAVSO International Database (https://www.aavso.org/aavso-international-database).
Kenyon, S. J. 1986, *The Symbiotic Stars*, Cambridge Univ. Press, Cambridge.
Lenz, P., and Breger, M. 2005, *Commun. Asteroseismology*, **146**, 53.
Pojmański, G. 2002, *Acta Astron.*, **52**, 397.
Renault, C., *et al.* 1998, *Astron. Astrophys.*, **329**, 522.
Schlafly, E. F., and Finkbeiner, D. P. 2011, *Astrophys. J.*, **737**, 103.
Skopal, A. 2008, *J. Amer. Assoc. Var. Star Obs.*, **36**, 9.
Snyder, L. F., and Lapham, J. 2008, in *The Society for Astronomical Sciences 27th Annual Symposium on Telescope Science, held May 20–22, 2008 at Big Bear Lake, CA,* Society for Astronomical Sciences, Rancho Cucamonga, CA, 29.
Zacharias, N., Finch, C. T., Girard, T. M., Henden, A., Bartlett, J. L., Monet, D. G., and Zacharias, M. I. 2012, The Fourth U.S. Naval Observatory CCD Astrograph Catalog (UCAC4), VizieR On-line Data Catalog (http://cdsarc.u-strasbg.fr/viz-bin/Cat?I/322).